\documentclass[%
 reprint,
superscriptaddress,
 amsmath,amssymb,
 aps,
pra,
]{revtex4-2}

\usepackage[colorlinks = true,
            linkcolor = blue,
            urlcolor  = black,
            citecolor = blue,
            anchorcolor = blue]{hyperref}
\usepackage{physics}
\usepackage{graphicx}
\usepackage{dcolumn}
\usepackage{bm}
\usepackage{float}

\begin{document}

\preprint{APS/123-QED}

\title{Environment-induced entanglement generation for two qubits in the presence of qubit-qubit interaction noise}

\author{Muhammad Abdullah Mutahar}
\affiliation{School of Science \& Engineering, Lahore University of Management Sciences (LUMS),\\
Opposite Sector U, D.H.A, Lahore 54792, Pakistan}
 
\author{Adam Zaman Chaudhry}
 \email{adam.zaman@lums.edu.pk}
\affiliation{School of Science \& Engineering, Lahore University of Management Sciences (LUMS),\\
Opposite Sector U, D.H.A, Lahore 54792, Pakistan}






\begin{abstract}
Using an exactly solvable pure dephasing model, we show how entanglement between qubits can be generated via the interaction with a common environment and concurrent application of suitable control pulses. The control pulses are able to effectively remove the detrimental effect of the environment while preserving the indirect interaction between the qubits, thereby leading to the generation of near-perfect entanglement. Furthermore, we also investigate the entanglement dynamics if the qubits are directly interacting; this interaction may even contain a noise term. The present of this additional noise leads to an additional decoherence term. This decoherence term cannot be removed by applying the pulses at the same time to both qubits. Rather, we show that by introducing a time delay between the two pulse sequences, near-perfect entanglement can still be generated via the interaction with the common environment.
 
\end{abstract}

\maketitle


\section{\label{sec:level1} Introduction}

Entanglement is a fundamental concept in quantum mechanics, and one of the key resources in quantum information processing \cite{nielsen2000chuang,bennett1992quantum,bennett1993teleporting,schumacher1998quantum}. It finds many practical applications in quantum computation and communication such as quantum cryptography \cite{ekert1992quantum,deutsch1996quantum} and superdense coding \cite{bennett1992communication}. Unfortunately, quantum resources such as entanglement and coherence are very delicate - realistic quantum systems inevitably interact with their surrounding environments, deteriorating these resources very quickly as a result of the decoherence process \cite{zurek2003decoherence,schlosshauer2007decoherence,breuer2002theory}. Consequently, the environment has largely been considered very problematic for quantum technologies. However, with the advent of ‘reservoir engineering', it has been shown that the system-environment interaction can instead be used to generate useful quantum resources \cite{zagoskin2006quantum,neeley2008process,murch2012cavity,cirac1993dark,poyatos1996quantum,carvalho2001decoherence}. This is carried out by tuning the system-environment coupling and properties of the environment. These techniques have been used experimentally to generate superposition states in superconducting circuits \cite{shankar2013autonomously}, atomic ensembles \cite{krauter2011entanglement}, and trapped ions \cite{barreiro2011open,lin2013dissipative}. \\

Of key interest to us in this work is the fact that two independent qubits interacting with a common environment can become entangled \cite{orszag2010coherence,PhysRevLett.89.277901,PhysRevA.77.062303}. This environment-induced entanglement has several applications, including, but not limited to, quantum control of two-dimensional quantum systems \cite{PhysRevLett.97.080402}, entanglement of two-mode squeezed states \cite{PhysRevA.76.042127,PhysRevA.79.032102}, and entanglement of charged qubits \cite{PhysRevB.77.155420}. Therefore, it is worthwhile to optimize and enhance this environment-induced entanglement. Attention has been directed towards optimising the properties of the environment to enhance the induced entanglement dynamics \cite{chaudhry2015optimization}. On the other hand, considerable work has been done to mitigate the detrimental influence of the environment. Inspired by the technique of dynamical decoupling that uses suitable control fields to remove the effect of the environment \cite{PhysRevA.89.024101,1PhysRevA.58.2733,2PhysRevLett.82.2417,3PhysRevLett.85.3520,4PhysRevLett.88.207902,5yang2011preserving,6PhysRevA.64.052301,7PhysRevLett.90.037901,8PhysRevA.73.062317,9PhysRevLett.101.010403,10PhysRevLett.102.080405,11PhysRevLett.98.100504,ChaudhryPRA12012,ChaudhryPRA22012}, in this paper we investigate the effect of pulse sequences on the entanglement dynamics of two qubits interacting with a common environment. Our hope is that the applied control pulses would be such that the decoherence of the two qubits is highly mitigated while largely leaving the entanglement generating influence of the environment untouched.\\

We start by considering a system of two independent qubits, initially
in a pure product state, interacting with a common environment. This environment is modeled as a
collection of harmonic oscillators. As the dephasing time is much shorter than the relaxation time scale, we restrict our model to the pure dephasing case. After working out the exact two-qubit dynamics in the presence of pulses, we show that near-perfect entanglement can be generated via the application of pulses. We then further modify our system to include a direct noisy interaction between the two qubits. We show that, as a result of the noise in the interaction between the qubits, an additional decoherence term emerges. Interestingly, to mitigate the influence of this additional term on the entanglement, the pulses cannot be applied at the same time to both qubits. Rather, there must be a time delay. By incorporating such a time-delay, we show how near-perfect entanglement can still be generated.

This paper is organised in the following way. In Sec.~\ref{without noise}, we introduce our basic model and work out its dynamics. In Sec.~\ref{generating_entanglement}, we work out the entanglement dynamics to show the near-perfect generation of entanglement with the application of pulses. Then, in Sec.~\ref{with noise}, we modify our Hamiltonian to accommodate for the qubits interacting directly with each other. This interaction term contains a classical stochastic variable depicting a noisy interaction. We conclude our findings in Sec.~\ref{conclusion}, with some details deferred to the appendix.

\section{The model and its dynamics} \label{without noise}

We begin with two qubits interacting with a common bosonic environment. The total system-environment Hamiltonian is then given by (we use $\hbar = 1$ throughout)
\begin{equation} \label{hamiltonain}
\begin{aligned}
H &= H_{S} + H_{B} + H_{I} + H_{C} \\
&= \frac{\omega_{0}}{2} \left( \sigma_{z}^{(1)} + \sigma_{z}^{(2)} \right) + \sum_{k}\omega_{k}b_{k}^{\dagger}b_{k} \\ 
&\hspace{4mm} + \left( \sigma_{z}^{(1)} + \sigma_{z}^{(2)} \right) \sum_{k} \left(g_{k}^{*}b_{k} + g_{k}b_{k}^{\dagger}\right)\\
&\hspace{4mm} + H_{C}^{(1)} + H_{C}^{(2)}.
\end{aligned}
\end{equation}
Here, the superscripts $(1)$ and $(2)$ refer to the first qubit and the second qubit, $\omega_{0}$ is the spacing between the energy states of a two-level system (assumed to be the same for both qubits), $\sigma_{z}$ is the standard Pauli matrix, while $H_{B} = \sum_{k}\omega_{k}b_{k}^{\dagger}b_{k} $ represents the common bosonic environment composed of a collection of harmonic oscillators. The third term describes the qubit-environment interaction for the qubits, while the last terms, $H_{C}^{(1)}$ and $H_{C}^{(2)}$, represent the control fields (pulses) applied to the first and the second qubit respectively. We note that our system Hamiltonian is restricted to the pure-dephasing case of the widely known spin-boson model for two qubits \cite{doi:10.1098/rspa.1996.0029,PhysRevA.65.032326}. This restriction is allowed if we consider the dynamics within a time scale much shorter than the relaxation
time-scale of the system.

To solve the dynamics, we first consider the free time-evolution operator $U_{0}(t) = \text{exp}[-i(H_{S}+H_{B})t]$. The Hamiltonian in the interaction picture is then defined as 
\begin{equation} \label{interaction Hamiltonian}
\begin{aligned}
H_{I}(t) &= U_{0}^{\dagger}(t) H_{I} U_{0}(t) + U_{0}^{\dagger}(t) H_{C}^{(1)} U_{0}(t)\\  &\hspace{4mm} + U_{0}^{\dagger}(t) H_{C}^{(2)} U_{0}(t)\\
&= \left( \sigma_{z}^{(1)} + \sigma_{z}^{(2)} \right) \sum_{k} \left(g_{k}^{*}b_{k} e^{-i \omega_{k} t} + g_{k}b_{k}^{\dagger} e^{i \omega_{k} t}\right)\\
&\hspace{4mm} +  H_{C}^{(1)}(t) + H_{C}^{(2)}(t).
\end{aligned}
\end{equation}
It is useful to move to the toggling frame of the pulses applied on the qubits.
Assuming that $H_{C}^{(1)}(t)$ and $H_{C}^{(2)}(t)$ generate a series of instantaneous $\pi$ rotations of the qubits around an axis orthogonal to the $z$-axis at times $t_{k}$, where $0 \leq k \leq n+1$ with $n$ being the number of pulses applied and  $t_{0}=0, t_{n+1}=t$, the interaction Hamiltonian in the toggling frame of the pulses takes the form
\begin{equation}
\begin{aligned}
H_{T}(t) &= \left( s^{(1)}(t)\sigma_{z}^{(1)} + Z^{(2)}(t)\sigma_{z}^{(2)} \right)\\
&\hspace{4mm} \times \sum_{k} \left(g_{k}^{*}b_{k} e^{-i \omega_{k} t} + g_{k}b_{k}^{\dagger} e^{i \omega_{k} t}\right),\\
\end{aligned}
\end{equation}
where
\begin{align*}
    Z^{(1)}(t) = \int_{0}^{t} s^{(1)}(t') dt' ,\\
    Z^{(2)}(t) = \int_{0}^{t} s^{(2)}(t') dt'.
\end{align*}
As the role of the pulses can essentially be reduced to flipping the sign of $\sigma_{z}$, we have attached a switching function $s(t)$ above for both the qubits, and the superscript is there to remind us that the applied pulse sequence need not be the same for both qubits. The switching function is
\begin{equation}
s(t') = \sum_{k=0}^{n} (-1)^{k} \Theta\left(t_{k+1}-t'\right) \Theta\left(t'-t_{k}\right)   , 
\end{equation}
where $\Theta(t)$ is the Heaviside step function. We further denote the times at which pulses are applied $t_{k}$ as $t_{k} = \delta_{k}t$ with $0\leq \delta_{k} \leq 1$, where $k=1,\hdots,n$. 
Next, we move on to evaluating the unitary time evolution operator corresponding to $H_{T}(t)$. This is found to be 
\begin{equation} \label{uni-time}
\begin{aligned}
U_{I}(t) &= \text{exp} \left[\sum_{i=1}^{2} A_{i}\right].
\end{aligned}
\end{equation}
Here 
\begin{equation} \label{A1}
\begin{aligned}
A_{1}(t) = &-i \sigma_{z}^{(1)} \sum_{k} \left[ \phi^{*(1)}_{k}(t) g_{k}^{*} b_{k} + \phi^{(1)}_{k}(t) g_{k} b_{k}^{\dagger}\right] \\
&-i \sigma_{z}^{(2)} \sum_{k} \left[ \phi^{*(2)}_{k}(t) g_{k}^{*} b_{k} + \phi^{(2)}_{k}(t) g_{k} b_{k}^{\dagger}\right], \\
= &-i \sigma_{z}^{(1)} \chi_{1}(t) - i \sigma_{z}^{(2)} \chi_{2}(t),
\end{aligned}
\end{equation}
where
\begin{equation} \label{phi}
\begin{aligned}
\phi_{k}^{(1)}(t) &\equiv \int_{0}^{t} dt' s^{(1)}(t') e^{i \omega_{k} t'}, \\
\phi_{k}^{(2)}(t) &\equiv \int_{0}^{t} dt' s^{(2)}(t') e^{i \omega_{k} t'}, \\
\end{aligned}
\end{equation}
and $\chi_1(t)$ and $\chi_2(t)$ are defined using Eq.~\eqref{A1}. We work out $A_2(t)$ to be 
\begin{equation} \label{A2}
\begin{aligned}
A_{2}(t) &= i B_{1}(t) + i B_{2}(t) + i B_{3}(t) \sigma_{z}^{(1)} \sigma_{z}^{(2)}\\
&\hspace{4mm} + i B_{4}(t) \sigma_{z}^{(1)} \sigma_{z}^{(2)},
\end{aligned}
\end{equation}
with
\begin{equation} \label{Bi}
\begin{aligned}
B_{1}(t) =& i \sum_{k} \left|g_{k}\right|^{2}\\ 
&\times \int_{0}^{t} dt_{1} \int_{0}^{t_{1}} dt_{2} s^{(1)}(t_{1}) s^{(1)}(t_{2}) \sin[\omega_{k}(t_{1}-t_{2})],\\
B_{2}(t) =& i \sum_{k} \left|g_{k}\right|^{2} \\
&\times \int_{0}^{t} dt_{1} \int_{0}^{t_{1}} dt_{2} s^{(2)}(t_{1}) s^{(2)}(t_{2}) \sin[\omega_{k}(t_{1}-t_{2})],\\
B_{3}(t) =& i \sum_{k} \left|g_{k}\right|^{2} \\
&\times \int_{0}^{t} dt_{1} \int_{0}^{t_{1}} dt_{2} s^{(1)}(t_{1}) s^{(2)}(t_{2}) \sin[\omega_{k}(t_{1}-t_{2})],\\
B_{4}(t) =& i \sum_{k} \left|g_{k}\right|^{2}\\
&\times \int_{0}^{t} dt_{1} \int_{0}^{t_{1}} dt_{2} s^{(2)}(t_{1}) s^{(1)}(t_{2}) \sin[\omega_{k}(t_{1}-t_{2})].
\end{aligned}
\end{equation}
The full unitary time evolution operator corresponding to the Hamiltonian in the toggling frame of the control fields then becomes
\begin{equation}
\begin{aligned}
U(t) &= \text{exp}\left[\frac{-i \omega_{0}}{2} \left(Z^{(1)}(t) \sigma_{z}^{(1)} + Z^{(2)}(t)\sigma_{z}^{(2)}\right)\right] \\
&\hspace{4mm}\times \text{exp}\left[  -i H_{B} t  \right] \\
&\hspace{4mm}\times \text{exp}\left[ -i \sigma_{z}^{(1)} \chi_{1}(t) \right]  \text{exp}\left[ -i \sigma_{z}^{(2)} \chi_{2}(t) \right] \\
&\hspace{4mm}\times \text{exp}\left[  i \left(B_{1}(t) + B_{2}(t)\right)  \right] \\
&\hspace{4mm}\times \text{exp}\left[  i \left(B_{3}(t) + B_{4}(t)\right) \sigma_{z}^{(1)}\sigma_{z}^{(2)} \right].
\end{aligned}
\end{equation}

We now proceed to compute the dynamics of the system in the density matrix formalism, given any arbitrary system-environment initial state. For simplicity we consider the initial state to be a product state of any qubit-system state with the environment in a thermal bath state
\begin{equation} \label{density state}
\begin{aligned}
    \rho(0) &= \rho_{S}(0) \otimes \rho_{B}, \\
    \rho_{B} &= \frac{\text{exp} \left[-\beta H_{B}\right]}{Z_{B}}, \\
    Z_{B} &= \text{Tr}_{B}[e^{-\beta H_{B}}].
\end{aligned}
\end{equation}
Here, $\beta$ is the inverse temperature and $Z_{B}$ is the partition function of the environment. $\text{Tr}_{B}[\hspace{2mm}]$ represents the trace over the environment states. Given the initial system-environment state, we can work out the state at a later time $t$ via $\rho(t) = U(t) \rho(0) U^{\dagger}(t)$. We do so in the basis $\ket{k,l}$ given by
\begin{equation*}
\begin{aligned}
\sigma_{z}^{(1)} \ket{k,l} &= k \ket{k,l}, \\
\sigma_{z}^{(2)} \ket{k,l} &= l \ket{k,l}.
\end{aligned}
\end{equation*}
It is also useful to define 
\begin{equation}
\begin{aligned}
\bra{k',l'}  \rho_{S}(t) \ket{k,l} &= \left[   \rho_{S}(t)     \right]_{k',l';k,l} = \text{Tr}_S [\rho_S(t) P_{kl,k'l'}],
\end{aligned}
\end{equation}
where $P_{kl,k'l'} = \ket{k,l} \bra{k',l'}$, and $\rho_S(t) = \text{Tr}_B [\rho(t)]$ is the density matrix of the two qubits only. Now, it is easy to show that $\left[\rho_{S}(t)\right]_{k',l';k,l} = \text{Tr}_{S+B} \left[  U^{\dagger}(t)  P_{kl,k'l'}   U(t) \rho(0)  \right]$, where $\text{Tr}_{S+B}$ is the total trace over both system and environment. Using the expression for the total time-evolution operator already worked out, we find that 
\begin{equation} \label{rho}
\begin{aligned}
\left[\rho_{S}(t)\right]_{k',l';k,l} &= \left[\rho_{S}(0)\right]_{k',l';k,l} \\
&\hspace{4mm} \times \text{exp}\left[\frac{i \omega_{0}}{2} \left(k \hspace{0.5mm} Z^{(1)}(t) + l \hspace{0.5mm} Z^{(2)}(t) \right)\right]\\
&\hspace{4mm} \times \text{exp}\left[\frac{-i \omega_{0}}{2} \left(k' \hspace{0.5mm} Z^{(1)}(t) + l' \hspace{0.5mm} Z^{(2)}(t)\right)\right]\\
&\hspace{4mm} \times \text{exp} \left[-i B(t) \left(kl-k'l'\right)\right] \\
&\hspace{4mm} \times \text{Tr}_{B}\left[e^{i R_{kl,k'l'} } \rho_{B}\right].
\end{aligned}
\end{equation}
Here $B(t) = B_{3}(t)+B_{4}(t)$ and $R_{kl,k'l'} = (k-k')\chi_{1}(t) + (l-l')\chi_{2}(t)$. To further evaluate the trace in the last line above, we use the identity $\text{Tr}_{B}[e^{A}\rho_{B}] = e^{\text{Tr}_{B}[A^2 \rho_{B}]/2}$, leading to
\begin{equation} \label{Tr}
    \text{Tr}_{B}\left[e^{i R_{kl,k'l'} } \rho_{B}\right] = \text{exp} \left[-\frac{1}{2} \text{Tr}_{B}\left[\left(R_{kl,k'l'}\right)^{2} \rho_{B}\right]\right],
\end{equation}
with
\begin{equation} \label{R}
\begin{aligned}
\text{Tr}_{B} & \left[\left(R_{kl,k'l'}\right)^{2} \rho_{B}\right] =\\
&(k-k')^2 \sum_{r} \left|\phi_{r}^{(1)}(t)\right|^2 \left|g_{r} \right|^2 \left(2n_{r}+1\right)\\
+&(l-l')^2 \sum_{r} \left|\phi_{r}^{(2)}(t)\right|^2 \left|g_{r} \right|^2 \left(2n_{r}+1\right)\\
+&2(k-k')(l-l') \sum_{r} \phi_{r}^{*(1)}(t) \hspace{1mm}\phi_{r}^{(2)}(t) \left|g_{r} \right|^2 \left(n_{r}+1\right)\\
+&2(k-k')(l-l') \sum_{r} \phi_{r}^{(1)}(t) \hspace{1mm} \phi_{r}^{*(2)}(t) \left|g_{r} \right|^2 n_{r}.
\end{aligned}
\end{equation}
Note that  $\text{Tr}_{B}\left[b_{r}^{\dagger}b_{r} \rho_{B}\right] = n_{r} = 1/(e^{\beta \omega_{r}}-1)$, which is Bose-Einstein distribution since the environment is in thermal equilibrium.

Until this point we have kept the formalism arbitrary with respect to the pulses. The purpose for keeping it arbitrary, as we would see later, is to use this formalism to analyze the more complicated problem where a `noisy' qubit-qubit interaction is present as well. However, for our current purpose, we now assume that both qubits are subjected to the same pulse sequence. This would imply $s^{(1)}(t) = s^{(2)}(t) = s(t)$, $Z^{(1)}(t) = Z^{(2)}(t)=Z(t)$,  $\phi_{r}^{(1)}(t) = \phi_{r}^{(2)}(t) = \phi_{r}(t)$ and $B_{3}(t) = B_{4}(t) = D(t)$. By this assumption, Eq.~\eqref{R} becomes 
\begin{multline*}
\text{Tr}_{B} \left[\left(R_{kl,k'l'}\right)^{2} \rho_{B}\right] =\\
\left(k+l-k'-l' \right)^2 \sum_{r} \left|\phi_{r}(t)\right|^2 \left|g_{r}\right|^2 \coth\left(\frac{\beta\omega_{r}}{2}\right).
\end{multline*}
It is useful to further express our formalism in terms of filter functions \cite{cywinski2008enhance}. The action of a pulse sequence is captured by its filter function which is essentially in terms of the Fourier transform of $s(t)$. The filter function is defined as 
\begin{equation} \label{filter function}
    F(\omega t) = \frac{\omega^2}{2} \left|\int_{0}^{t} s(t') e^{i\omega t'} dt' \right|^2.
\end{equation}
At this point, the spectral density $J(\omega)$ is also introduced via the substitution 
\begin{equation} \label{spectral density}
    \sum_{r} 4 \left| g_{r}  \right|^2 \rightarrow \int_{0}^{\infty} d\omega J(\omega)\delta\left(\omega_{r}-\omega\right).
\end{equation}
This replaces the discrete summation with a continuous spectrum of environment frequencies $\omega$. The spectral density $J(\omega)$ is typically expressed using a power law, i.e $J(\omega) = g \frac{\omega^s}{\omega_{c}^{s-1}}G(\omega,\omega_{c})$, where $g$ is the system-environment coupling strength, $s$ is the Ohmicity parameter, and $G(\omega,\omega_{c})$ is the cutoff function with $\omega_{c}$ being the cutoff frequency. For the sake of simplicity, we choose to work in the Ohmic ($s=1$) regime \cite{weiss2012quantum,breuer2002theory}. Similarly, although different varieties of cutoff functions are used in literature \cite{breuer2002theory}, for simplicity, we would stick to the exponential cutoff function, that is, $G(\omega,\omega_{c}) = \text{exp}\left[-\omega/\omega_{c}\right]$.\\

In summary, putting everything together, we finally have
\begin{equation} \label{final rho}
\begin{aligned}
\left[\rho_{S}(t)\right]_{k',l';k,l} &= \left[\rho_{S}(0)\right]_{k',l';k,l} \\
&\hspace{4mm} \times \text{exp}\left[\frac{-i \omega_{0} Z(t)}{2} \left(k'+l'-k-l \right) \right]\\
&\hspace{4mm} \times \text{exp} \left[ \frac{i D(t)}{2} \left(k'l'-kl\right)\right] \\
&\hspace{4mm} \times \text{exp} \left[  -\frac{\gamma(t)}{4} \left(k+l-k'-l'\right)^2           \right],
\end{aligned}
\end{equation}
where 
\begin{align*}
    D(t) &= \int_{0}^{\infty}d\omega J(\omega) \int_{0}^{t}dt_{1} \hspace{1mm} s(t_{1}) \\ &\times \int_{0}^{t_{1}}dt_{2} \hspace{1mm} s(t_{2}) \sin\left[\omega\left(t_{1} - t_{2}\right)\right],\\
    \gamma(t) &= \int_{0}^{\infty} d\omega J(\omega) \frac{F\left(\omega t\right)}{\omega^2} \coth\left(\frac{\beta\omega}{2} \right)   . 
\end{align*}

\section{Generating entanglement} \label{generating_entanglement}

As stated in the introduction, our goal is to analyze the entanglement generation of two qubits under various pulse sequences and evaluate if the pulse sequences can possibly enhance the entanglement generation of the two qubits as compared to the qubits interacting with the common environment without any pulses applied. Let us try to interpret Eq.~\eqref{final rho}. The first factor containing $\text{exp}\left[-i\omega_{0} Z(t)\left(k'+l'-k-l \right)/2 \right]$ describes the free evolution of the qubits in the presence of the pulses. The second factor, that is, containing $\text{exp} \left[ \frac{i D(t)}{2} \left(k'l'-kl\right)\right]$ describes the indirect interaction of the qubits, modulated by the pulses, due to the common bosonic environment. It is this interaction that generates the entanglement, so our goal would be to preserve this interaction as much as we could. The third term, namely $\left[  -\frac{\gamma(t)}{4} \left(k+l-k'-l'\right)^2           \right]$, represents the decoherence factor $\gamma(t)$. We would try to minimize this term as much as we can by applying control pulses with short pulse spacing - this, of course, is the idea behind dynamical decoupling. Essentially, the decoherence rate depends on the overlap between the spectral density and the filter function; if pulses are applied rapidly, this overlap becomes very small, leading to a very small decoherence rate. However, the indirect interaction can still survive.

Let us then analyze the functions $D(t)$ and $\gamma(t)$ in more detail. Starting with $D(t)$, 
\begin{align*}
    D(t) = \int_{0}^{\infty}d\omega \hspace{1mm} J(\omega) \hspace{1mm} d(\omega,t),
\end{align*}
where
\begin{align*}
    d(\omega,t) = \int_{0}^{t}dt_{1} \int_{0}^{t_{1}}dt_{2} \hspace{1mm} s(t_{1}) s(t_{2}) \sin\left[\omega\left(t_{1} - t_{2}\right)\right].
\end{align*}
For $n$ pulses, we denote $d(\omega,t)$ by $d_{n}(\omega,t)$, which can be simplified to \cite{PhysRevA.89.063604}
\begin{align*}
    d_{n}(\omega,t) = \vartheta(\omega,t) + \nu(\omega,t) - t/\omega,
\end{align*}
with
\begin{align*}
    \vartheta(\omega,t) =& \frac{1}{\omega^2} \left[ 2 \sum_{m=1}^{n} (-1)^{m} \sin(\omega t_{m}) + (-1)^{n+2} \sin(\omega t) \right], \\
    \nu(\omega, t) =& \frac{2}{\omega^2} \Biggl\{   \sum_{m=1}^{n} \sum_{j=1}^{n} (-1)^{m+j} \left(\sin\left[\omega \left(t_{m+1} - t_{j}\right)\right]\right)     \\
    &- \sin \left[ \omega \left(t_{m}  - t_{j}\right)\right] \Biggl\}.
\end{align*}
For $\vartheta(\omega,t)$, if we assume a very large number of pulses ($n>>1$), we see that the first term effectively approaches zero since it is rapidly switching sign, and, in the high pulse limit, $t_{m + 1}$ approaches $\approx t_{m}$. The same logic applies to $\nu(\omega,t)$. We are then left with a very simple $d(\omega,t)$ in the high pulse limit. Coming towards $\gamma(t)$, it is useful to first define $F_{1}(\omega) = J(\omega)\coth(\beta\omega/2)$, and $F_{2}(\omega)=F(\omega t_{0})/ \omega^2$. We can then rewrite $\gamma(t)$ as 
\begin{align} \label{gamma}
    \gamma(t_{0}) = \int_{0}^{\infty} d\omega F_{1}(\omega) F_{2}(\omega).
\end{align}
It can be seen that the maximum contribution to the integral occurs where the functions $F_{1}(\omega)$ and $F_{2}(\omega)$ overlap the most. The effect of increasing the number of pulses $n$ can essentially be understood as the shifting of the peak of the filter function and hence of $F_{2}(\omega)$ towards the side of increasing $\omega$. As $F_{1}(\omega)$ is an exponentially decreasing function after $\omega = \omega_{c}$, the overlap of both functions keeps on decreasing as we increase the number of pulses. This also suggests in the high pulse limit ($n >> 1$), $\gamma(t) \rightarrow 0$. Therefore, we can expect to get negligible decoherence but significant entanglement due to the indirect interaction encapsulated by $D(t)$.

\begin{figure}[t]
\includegraphics[width=\linewidth]{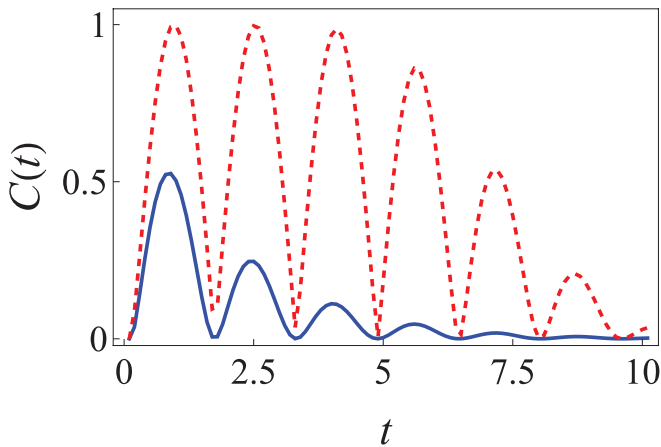}
\caption{\label{fig:C_freevspulses} Concurrence $C(t)$ in the absence of pulses vs pulses applied according to the PDD scheme for $n=256$. The solid, blue line represents the concurrence in the absence of pulses, while the red, dashed line represents the $C(t)$ with the applied pulses. We are using dimensionless units throughout with $\hbar = 1$, and we have used $\omega_{c}=20$, $\beta=1$, and $g=0.1$.}
\end{figure}

We now numerically check our claim regarding the generation of entanglement. First, to compute the dynamics of entanglement between the two qubits, we the use concurrence $C(t)$ \cite{wootters1998entanglement}. It is defined via the Hermitian matrix $V$ such that
\begin{align*}
V = \sqrt{\sqrt{\rho_{S}}\hspace{1mm}\widetilde{\rho_{S}}\hspace{1mm}\sqrt{\rho_{S}}},   
\end{align*}
where
\begin{align*}
\widetilde{\rho_{S}} = \left(\sigma_{y} \otimes \sigma_{y}\right) \rho_{S}^{*} \left(\sigma_{y} \otimes \sigma_{y}\right).
\end{align*}
Here, $\sigma_{y}$ is the standard Pauli matrix. $C(\rho_{S})$ is then computed via
\begin{equation} \label{C(t)}
C(\rho_{S}) = \text{max}\left[0, \lambda_{1} - \lambda_{2} - \lambda_{3} - \lambda_{4} \right].
\end{equation}
The $\lambda_{i}$ are the eigenvalues of $V$ in decreasing order. Now, in order to see the dynamics of entanglement, we should start with the initial system state to be a product state of the two qubits, and then we would analyse if and how the environment generates entanglement in between those qubits. Furthermore, the initial two-qubit state should not be part of a decoherence-free subspace. We begin our analysis by choosing the initial system state to be $\rho_{S}(t) = \ket{+,+}\bra{+,+}$ where $\sigma_{x}\ket{+} = \ket{+}$. The system density matrix at time $t$ is then given by
\begin{widetext}
\begin{equation}
  \rho_{S}(t) = \frac{1}{4}
  \begin{pmatrix}
    1 & e^{-i\omega_{0}Z(t)}e^{-\gamma(t)+iD(t)} & e^{-i\omega_{0}Z(t)}e^{-\gamma(t)+iD(t)} & e^{-4\gamma(t)-2i\omega_{0}Z(t)} \\
    e^{i\omega_{0}Z(t)}e^{-\gamma(t)-iD(t)} & 1 & 1 & e^{-i\omega_{0}Z(t)}e^{-\gamma(t)-iD(t)} \\
    e^{i\omega_{0}Z(t)}e^{-\gamma(t)-iD(t)} & 1 & 1 & e^{-i\omega_{0}Z(t)}e^{-\gamma(t)-iD(t)} \\
    e^{-4\gamma(t)+2i\omega_{0}Z(t)} & e^{i\omega_{0}Z(t)}e^{-\gamma(t)+iD(t)} & e^{i\omega_{0}Z(t)}e^{-\gamma(t)+iD(t)} & 1
  \end{pmatrix}.
  \label{eq:myeq}
\end{equation}
\end{widetext}
We now investigate concurrence $C(t)$ against time $t$ by applying pulses according to the periodic dynamical decoupling (PDD) scheme to both qubits, and compare it with the evolution when no pulses are applied. Results are illustrated in Fig.~\ref{fig:C_freevspulses}. By comparing the solid, blue line with the red, dashed line, it is clear that near-perfect entanglement can be generated via the application of the pulses. These results are in accordance with our prediction that the effect of the decoherence factor is greatly nullified, while the indirect interaction remains. Note, however, that after some time, the decoherence becomes significant, leading to a decrease in the value of the peak entanglement.

\begin{figure}[ht]
\includegraphics[width=\linewidth]{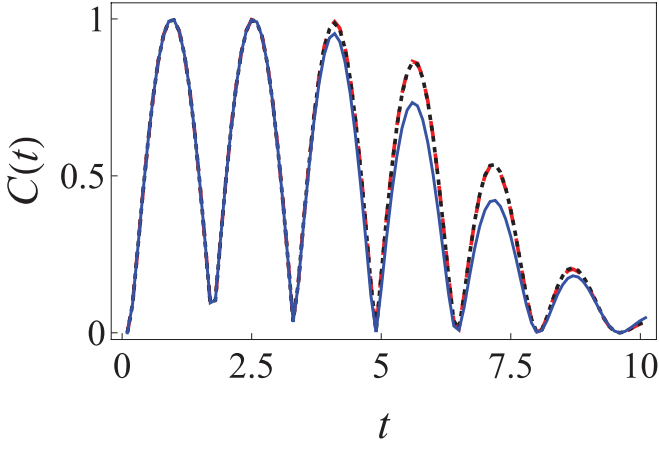}
\caption{\label{fig:PDDvsCPMGvsUDD} Concurrence $C(t)$ for PDD, CPMG and UDD pulses. The dashed, red line represents $C(t)$ for PDD pulses, the dotted, black line is for CPMG pulses, and the blue, solid line represents $C(t)$ for UDD pulses. The number of pulses applied is $n=256$. We use the same parameters as before, that is, $\omega_{c}=20$, $\beta=1$, and $g=0.1$.}
\end{figure}

We next consider the performance of different pulse sequences in generating entanglement. We numerically analyse $C(t)$ and compare the results for different types of pulse sequences, namely, Periodic Dynamical Decoupling (PDD) pulses, Carr-Purcell-Meiboom-Gill (CPMG) pulses\cite{carr1954effects,meiboom1958modified},  and Uhrig's dynamical decoupling (UDD) pulses \cite{uhrig2007keeping}. Results are shown in Fig.~\ref{fig:PDDvsCPMGvsUDD}. It is clear that while all three pulse sequences are able to generate entanglement, the entanglement decays in a different manner for the different sequences. This can be traced back to the fact that the mitigation of the decoherence factor with different pulse sequences is different.  

\section{Introducing the direct qubit-qubit interaction} \label{with noise}
\subsection{The Model and Time Evolution}

We now allow the two qubits to interact directly with each other. The system-environment Hamiltonian now becomes 
\begin{equation} \label{hamiltonain with interaction}
\begin{aligned}
H &= H_{S} + H_{B} + H_{I} + H_{C} \\
&= \frac{\omega_{0}}{2} \left( \sigma_{z}^{(1)} + \sigma_{z}^{(2)} \right) + \sum_{k}\omega_{k}b_{k}^{\dagger}b_{k} \\ 
&\hspace{4mm} + \lambda(t) \hspace{1mm} \sigma_{z}^{(1)} \sigma_{z}^{(2)} \\
&\hspace{4mm} + \left( \sigma_{z}^{(1)} + \sigma_{z}^{(2)} \right) \sum_{k} \left(g_{k}^{*}b_{k} + g_{k}b_{k}^{\dagger}\right)\\
&\hspace{4mm} + H_{C}^{(1)} + H_{C}^{(2)}
\end{aligned}
\end{equation}
where 
\begin{align*}
    \lambda(t) = \lambda_{0} + \xi(t)
\end{align*}
Here, $\lambda(t)$ represents the direct qubit-qubit interaction, expressed as the sum of a constant value $\lambda_0$ and a classical stochastic function $\xi(t)$. In other words, we allow for the possibility of noise in the interaction. For convenience, it is useful to express $\xi(t)$ in terms of a coupling constant $v$ and a dimensionless variable $\eta(t)$, that is, $\xi(t) = v \eta(t)$. We proceed as before and obtain the effective Hamiltonian in the toggling frame as 
\begin{equation}
\begin{aligned}
H_{T}(t) &= \lambda(t)\hspace{1mm} s^{(1)}(t) \hspace{1mm} s^{(2)}(t) \hspace{1mm} \sigma_{z}^{(1)} \sigma_{z}^{(2)} \\
&\hspace{4mm} + \left( s^{(1)}(t)\sigma_{z}^{(1)} + s^{(2)}(t)\sigma_{z}^{(2)} \right)\\
&\hspace{4mm} \times \sum_{k} \left(g_{k}^{*}b_{k} e^{-i \omega_{k} t} + g_{k}b_{k}^{\dagger} e^{i \omega_{k} t}\right)\\
\end{aligned}
\end{equation}
The unitary time-evolution operator corresponding to this is $U_{I}(t)$. Following our previous derivation, we now find that 
\begin{equation}
\begin{aligned}
A_{1}(t) = &-i \sigma_{z}^{(1)} \chi_{1}(t) - i \sigma_{z}^{(2)} \chi_{2}(t) \\
&-i \sigma_{z}^{(1)} \sigma_{z}^{(2)} \lambda_{0} h(t) -i \sigma_{z}^{(1)} \sigma_{z}^{(2)} \Gamma(t),
\end{aligned}
\end{equation}
where
\begin{align*}
    h(t) &= \int_{0}^{t} dt_{1} \hspace{1mm} s^{(1)}(t_{1}) \hspace{1mm} s^{(2)}(t_{1}), \\
    \Gamma(t) &= \int_{0}^{t} dt_{1} \hspace{1mm} \xi(t_{1}) \hspace{1mm} s^{(1)}(t_{1}) \hspace{1mm} s^{(2)}(t_{1}).
\end{align*}
Here $\chi_{1}(t)$ and $\chi_{2}(t)$ are given in Eq.~\eqref{A1}. $A_{2}(t)$ turns out to be same as Eq.~\eqref{A2} with $B_{i}(t)$ given in Eq.~\eqref{Bi}. With these changes, it is clear that the system state is now given by 
\begin{equation} \label{rho-interaction}
\begin{aligned}
\left[\rho_{S}(t)\right]_{k',l';k,l} &= \left[\rho_{S}(0)\right]_{k',l';k,l} \\
&\hspace{4mm} \times \text{exp}\left[\frac{i \omega_{0}}{2} \left(k \hspace{0.5mm} Z^{(1)}(t) + l \hspace{0.5mm} Z^{(2)}(t) \right)\right]\\
&\hspace{4mm} \times \text{exp}\left[\frac{-i \omega_{0}}{2} \left(k' \hspace{0.5mm} Z^{(1)}(t) + l' \hspace{0.5mm} Z^{(2)}(t)\right)\right]\\
&\hspace{4mm} \times \text{exp} \left[i B(t) \left(k'l'-kl\right)\right] \\
&\hspace{4mm} \times \text{exp} \left[  -i \lambda_{0} h(t) \left(k'l'-kl\right)  \right] \\
&\hspace{4mm} \times \text{exp} \left[  -i \Gamma(t) \left(k'l'-kl\right)  \right] \\
&\hspace{4mm} \times \text{Tr}_{B}\left[e^{i R_{kl,k'l'} } \rho_{B}\right]
\end{aligned}
\end{equation}
Here, $B(t)=B_{3}(t)+B_{4}(t)$ and $R_{kl,k'l'}=(k-k')\chi_{1}(t)+(l-l')\chi_{2}(t)$ as defined previously, and $\text{Tr}_{B}[e^{i R_{kl,k'l'} } \rho_{B}]$ is evaluated in Eq.~\eqref{Tr} and Eq.~\eqref{R}. \\

We now clearly observe the effect of the direct interaction in the terms $\text{exp} \left[  -i \lambda_{0} h(t) \left(k'l'-kl\right)  \right]$ and $\left[  -i \Gamma(t) \left(k'l'-kl\right)  \right]$. The former describes the effect of the constant interaction $\lambda_0$, while the latter is the effect of the noise in the interaction. Let us focus on the latter. Noticing the form of $\Gamma(t)$, it contains an integral over the stochastic function $\xi(t)$. To proceed, we compute $\Bigl \langle \text{exp} \left[  -i \Gamma(t) \left(k'l'-kl\right)  \right] \Bigr \rangle $, where $\langle ... \rangle$ corresponds to average over all possible noise realizations. We assume that the noise $\xi(t)$ is Gaussian, and that the noise is stationary. We then find that 
\begin{align*}
  \biggl \langle  e^{-i \Gamma(t) \left(k'l'-kl\right)} \biggr \rangle = e^{-\frac{1}{2}\left(k'l'-kl\right)^2 \bigl\langle \Gamma(t)^2 \bigl\rangle },
\end{align*}
with
\begin{align*}
    \biggl\langle \Gamma(t)^2 \biggl\rangle \hspace{1mm} =& \hspace{1mm} \biggl \langle \int_{0}^{t} dt_{1} \xi(t_{1}) s^{(1)}(t_{1}) s^{(2)}(t_{1}) \\
    & \times \int_{0}^{t} dt_{2} \xi(t_{2}) s^{(1)}(t_{2}) s^{(2)}(t_{2}) \biggl\rangle \\
    =& \int_{0}^{t} dt_{1} \int_{0}^{t} dt_{2} \hspace{1mm} s^{(1)}(t_{1}) s^{(2)}(t_{1}) s^{(1)}(t_{2}) s^{(2)}(t_{2}) \\
    & \times \biggl\langle \xi(t_{1}) \xi(t_{2}) \biggl\rangle 
\end{align*}
Introducing the change of variables $\tau = t_{1}-t_{2}$, $\bigl\langle \xi(t_{1}) \xi(t_{2}) \bigl\rangle = \bigl\langle \xi(\tau) \xi(0) \bigl\rangle$. We further write this autocorrelation function as the Fourier transform over the spectral density $S(\omega)$ of the noise, that is, $\bigl\langle \xi(\tau) \xi(0) \bigl\rangle = \frac{1}{2\pi} \int_{-\infty}^{\infty} d\omega\hspace{1mm} e^{-i \omega \tau}S(\omega)$. For simplicity, we introduce the combined switching function $f_{c}(t) = s^{(1)}(t) s^{(2)}(t)$. This allows us to write the Fourier transform of $\bigl\langle \Gamma(t)^2 \bigl\rangle$ as
\begin{equation} \label{mu}
\begin{aligned}
\biggl\langle \Gamma(t)^2 \biggl\rangle \hspace{1mm} &= \hspace{1mm} \frac{1}{2\pi} \int_{-\infty}^{\infty} d\omega \left|\int_{0}^{t} dt_{1} f_{c}(t_{1}) e^{-i\omega t_{1}}\right|^2 S(\omega) \\
&= \hspace{1mm} \frac{1}{\pi} \int_{-\infty}^{\infty} d\omega \frac{F_{c}(\omega t)}{\omega^2} S(\omega) \\
&= \hspace{1mm} \mu(t).
\end{aligned}
\end{equation}
Here $F_{c}(\omega t)$ is the filter function corresponding to $f_{c}(t)$.

The key point is that the presence of the noise in the interaction leads to the decay of the off-diagonal terms of the system density matrix. In other words, the noise in the interaction acts as an additional source of decoherence on top of the effect of $\gamma(t)$. Therefore, our goal now would be to eliminate the effect of both $\gamma(t)$ and $\mu(t)$, while preserving the effect of $D(t)$ and $\lambda_0$. We begin by noticing the form of $f_{c}(t)$. This is the product of the switching functions of the pulses applied to both qubits. If the same pulse sequence is used for both qubits, $f_{c}(t)=1$ regardless of the form of the pulse sequence because $s(t)^2 = 1$. This means that $\mu(t)$ would be unchanged upon the application of the pulses. Therefore, we investigate displacing the switching function of the pulse sequence on the second qubit with respect to the first qubit, that is, we consider $s^{(2)}(t) = s^{(1)}(t + \alpha)$, where $\alpha$ encapsulates the time shift between the pulse sequences applied to the two qubits. An important point to notice here is that when $\alpha = \Delta / 2$ for switching functions with constant pulse spacing $\Delta$, $f_{c}(t) = s^{(1)}(t) s^{(1)}(t+ \Delta/2)$ actually corresponds to a new switching function with double the number of pulses $n$. That is, if  $s^{(1)}(t)$ is a PDD sequence with $n$ pulses, $f_{c}(t)$ would be a PDD sequence with $2n$ pulses when $\alpha = \Delta/2$. This would make the computation of $F_{c}(\omega t)$ much simpler in this case. However, for pulse sequences with changing pulse spacing such as the UDD pulse sequence, we would have to compute the filter function for $f_{c}(t)$ for arbitrary values of $\alpha$ by Eq.~\eqref{filter function}. \\

We now note that, with $s^{(2)}(t) = s^{(1)}(t + \alpha)$, $\phi_{r}^{(2)}(t)$ is simplified to (see the appendix for details)
\begin{equation} \label{simplify}
    \phi_{r}^{(2)}(t) = e^{-i\omega_{r}\alpha} \phi_{r}^{(1)}(t).
\end{equation}
This greatly simplifies Eq.~\eqref{R}, which yields
\begin{equation}
\begin{aligned}
\text{Tr}_{B} & \left[\left(R_{kl,k'l'}\right)^{2} \rho_{B}\right] = \frac{1}{2}(k-k')^2 \gamma(t) + \frac{1}{2}(l-l')^2 \gamma(t)\\
+& (k-k')(l-l') \Bigl[2p(t) + q(t) + i r(t)\Bigl],
\end{aligned}
\end{equation}
where
\begin{align*}
    \gamma(t) &= \int_{0}^{\infty}d\omega \hspace{1mm} J(\omega) \frac{F(\omega t)}{\omega^2}\hspace{1mm} \text{coth}\left( \frac{\beta\omega}{2}   \right),\\
    p(t) &= \int_{0}^{\infty}d\omega \hspace{1mm} J(\omega) \frac{F(\omega t)}{\omega^2}\hspace{1mm} \text{cos}\left(\omega \alpha\right) \frac{1}{e^{\beta \omega}-1}, \\
    q(t) &= \int_{0}^{\infty}d\omega \hspace{1mm} J(\omega) \frac{F(\omega t)}{\omega^2}\hspace{1mm} \text{cos}\left(\omega \alpha\right), \\
    r(t) &= \int_{0}^{\infty}d\omega \hspace{1mm} J(\omega) \frac{F(\omega t)}{\omega^2}\hspace{1mm} \text{sin}\left(\omega \alpha\right).
\end{align*}
Here we have lifted the superscripts because everything is expressed in terms of $F^{(1)}(\omega t)$ using Eq.~\eqref{simplify}. This results in the density matrix $\rho_{S}(t)$ attaining the form
\begin{equation} 
\begin{aligned}
\left[\rho_{S}(t)\right]_{k',l';k,l} &= \left[\rho_{S}(0)\right]_{k',l';k,l} \\
&\hspace{4mm} \times \text{exp}\left[\frac{-i \omega_{0} Z(t)}{2} \left(k'+l'-k-l \right) \right]\\
&\hspace{4mm} \times \text{exp} \left[i B(t) \left(k'l'-kl\right)\right] \\
&\hspace{4mm} \times \text{exp} \left[  -i \lambda_{0} h(t) \left(k'l'-kl\right)  \right] \\
&\hspace{4mm} \times \text{exp} \left[ - \frac{1}{2} \left(k'l'-kl\right)^2 \mu(t) \right] \\
&\hspace{4mm} \times \text{exp}  \left[ -\frac{1}{4} (k'-k)^2 \gamma(t) \right] \\
&\hspace{4mm} \times \text{exp}  \left[ -\frac{1}{4} (l'-l)^2 \gamma(t) \right] \\
&\hspace{4mm} \times \text{exp}  \left[ -\frac{1}{2} (k'-k)(l'-l) \left[2p(t) + q(t)\right] \right] \\
&\hspace{4mm} \times \text{exp}  \left[ -\frac{i}{2} (k'-k)(l'-l) r(t)\right]
\end{aligned}
\end{equation}
To numerically study the dynamics of the two-qubit system, we need to specify the spectral density of the noise, which we denote by $S(\omega)$. For $1/f^a$ noise \cite{paladino20141}, this is given by 
\begin{equation}
S_{f}(\omega) = \frac{A_{0}^{1+a}}{\omega^{a}},    
\end{equation}
where we consider $0.5\leq a \leq 1.5$. As before, we choose the initial system state to be $\rho_{S}(t) = \ket{+,+}\bra{+,+}$. The system density matrix at time $t$ in matrix form is then given by
\begin{widetext}

\begin{equation}
  \rho_{S}(t) = \frac{1}{4}
  \begin{pmatrix}
    1 & e^{-i\omega_{0}Z(t)+iM(t)}e^{-\gamma(t)-2\mu(t)} & e^{-i\omega_{0}t+iM(t)}e^{-\gamma(t)-2\mu(t)} & e^{-2i\omega_{0}Z(t)-2ir(t)}e^{-2\gamma(t)-2R(t)} \\
    e^{i\omega_{0}Z(t)-iM(t)}e^{-\gamma(t)-2\mu(t)} & 1 & e^{-2\gamma(t)+2R(t)}e^{2ir(t)} & e^{-i\omega_{0}Z(t)-iM(t)}e^{-\gamma(t)-2\mu(t)} \\
    e^{i\omega_{0}Z(t)-iM(t)}e^{-\gamma(t)-2\mu(t)} & e^{-2\gamma(t)+2R(t)}e^{2ir(t)} & 1 & e^{-i\omega_{0}Z(t)-iM(t)}e^{-\gamma(t)-2\mu(t)} \\
    e^{2i\omega_{0}Z(t)-2ir(t)}e^{-2\gamma(t)-2R(t)} & e^{i\omega_{0}Z(t)+iM(t)}e^{-\gamma(t)-2\mu(t)} & e^{i\omega_{0}Z(t)+iM(t)}e^{-\gamma(t)-2\mu(t)} & 1
  \end{pmatrix}.
  \label{eq:myeqn}
\end{equation} \\
\end{widetext}
where $M(t) = 2B(t) - 2\lambda_{0}h(t)$ , and $R(t) = 2p(t)+q(t)$.

\begin{figure}[ht]
\includegraphics[width=\linewidth]{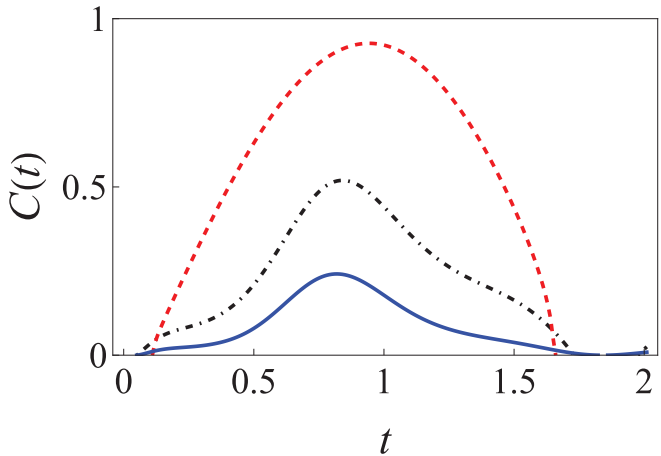}
\caption{\label{fig:Sf_freevsPDDvsPDDF} Concurrence $C(t)$ for the noise spectral density $S_{f}(\omega)$ with no pulses and in the presence of PDD pulses. The blue, solid line represents the concurrence with no pulses. The black, dot-dashed line is the concurrence with PDD pulses ($n = 371$) with zero time delay ($\alpha = 0$). The red, dashed line represents the concurrence of PDD pulse sequence for $n=371$ pulses with non-zero $\alpha$. The parameter used are $\alpha=0.0139$, $\omega_{ir}=2\pi$, $A_{0}=10$, $\lambda_{0}=0.1$, $\omega_{c}=20$, $\beta=1$, and $g=0.1$.}
\end{figure}

\begin{figure}[ht]
\includegraphics[width=\linewidth]{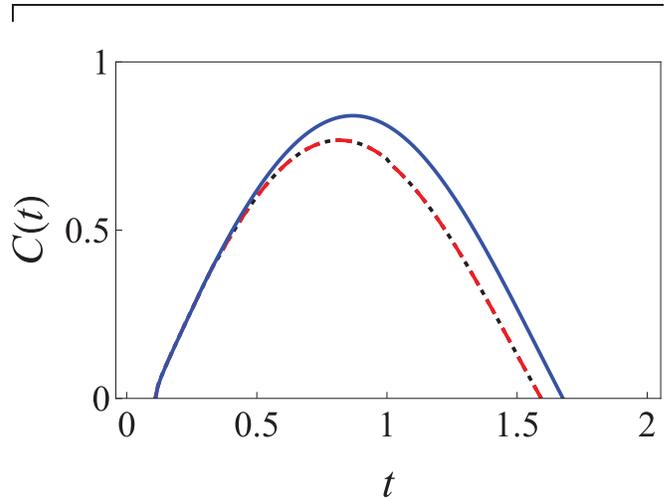}
\caption{\label{fig:Sf_PDDvsCDDvsUDD} Concurrence $C(t)$ with the noise spectral density $S_{f}(\omega)$ for PDD (black, short-dashed curve), CPMG (red, long-dashed curve), and UDD pulses (blue, solid curve). The number of pulses $n=64$ is kept constant for all pulse sequences. The parameters used are $\alpha=0.0139$, $\omega_{ir}=2\pi$, $A_{0}=10$, $\lambda_{0}=0.1$, $\omega_{c}=20$, $\beta=1$, and $g=0.1$.}
\end{figure}

We now numerically investigate the generation of entanglement. In Fig.~\ref{fig:Sf_freevsPDDvsPDDF}, we present the entanglement dynamics of the two qubits with and without PDD pulses. It is clear from the blue, solid curve that in the absence of any applied pulses, the decoherence factors from both the environment and the qubit-qubit interaction noise do not allow the environment-induced entanglement to be generated properly. Moreover, if there is no time-shift between the pulse sequences (that is, $\alpha = 0$), the system is able to overcome the decoherence factor of the environment, but the decoherence factor of the interaction noise $\mu(t)$ still leads to significant decoherence (see the black, dot-dashed curve). Finally, if we displace the switching function of the second qubit, that is, $s^{(2)}(t)=s^{(1)}(t+\alpha)$, $\mu(t)$ can also be minimised and near perfect entanglement can be generated. In Fig.~\ref{fig:Sf_PDDvsCDDvsUDD}, we illustrate that CPMG pulses and UDD pulses also generate significant entanglement when there is a time delay between the pulses applied to the two qubits.

\section{Conclusion} \label{conclusion}
We started with a two-qubit system interacting with a common bosonic environment and developed the formalism to compute the dynamics of the two qubits without any direct interaction. The results showed that for this model, applying control pulses to the two qubits does enhance the environment-generated entanglement between them. We then proceeded to add a qubit-qubit interaction term in the model Hamiltonian along with noise in this interaction. Due to this noise, the qubits were shown to undergo additional decoherence. To mitigate the effect of this additional term, it was shown that the same pulse sequence cannot be applied to both qubits, that is, there must be a time shift. If such a time shift is allowed, we showed that near perfect entanglement can again be generated. Our results should be useful in the context of reservoir engineering to generate entanglement between two qubits.


\appendix

\section{Simplifying $\phi_{r}^{(2)}(t)$} \label{phi-2}

Under the assumption taken in the paper that $s^{(2)}(t) = s^{(1)}(t+\alpha)$, $\phi_{r}^{(2)}(t)$ can be simplified as follows:

\begin{align*}
    \phi_{r}^{(2)}(t) &= \int_{0}^{t}dt' s^{(2)}(t') e^{i\omega_{r}t'} \\
    &=\int_{0}^{t}dt' s^{(1)}(t'+\alpha) e^{i\omega_{r}t'} \\
    &= e^{-i\omega_{r}\alpha} e^{i\omega_{r}\alpha}  \int_{0}^{t}dt' s^{(1)}(t'+\alpha) e^{i\omega_{r}t'} \\
    &= e^{-i\omega_{r}\alpha} \int_{0}^{t}dt' s^{(1)}(t'+\alpha) e^{i\omega_{r}(t'+\alpha)}. 
\end{align*}
Now performing a change of variables $t'' = t' + \alpha$, we get 
\begin{align*}
     \phi_{r}^{(2)}(t) =& e^{-i\omega_{r}\alpha} \int_{\alpha}^{t+\alpha}dt'' s^{(1)}(t'') e^{i\omega_{r}t''} \\
     =& e^{-i\omega_{r}\alpha} \Biggr[ \int_{0}^{t+\alpha}dt'' s^{(1)}(t'') e^{i\omega_{r}t''}\\
     &- \int_{0}^{\alpha}dt'' s^{(1)}(t'') e^{i\omega_{r}t''} \Biggr] \\
     =& e^{-i\omega_{r}\alpha} \left[ \phi_{r}^{(1)}(t+\alpha) - \phi_{r}^{(1)}(\alpha) \right].
\end{align*}
It can be shown that 
\begin{align*}
    F(\omega t) = \frac{1}{2}   \left| \sum_{k=0}^{n} (-1)^{k} \left(    e^{i\omega t_{k+1}} - e^{i\omega t_{k}}     \right)      \right|^2.
\end{align*}
Equating this with our definition of $\phi_{r}^{(1)}(t)$ in Eq.~\eqref{phi} and using Eq.~\eqref{filter function}, we get 
\begin{align*}
    F^{(1)}(\omega_{r} t) = \frac{\omega_{r}^{2}}{2}   \left| \phi_{r}^{(1)}(t) \right|^2.
\end{align*}
This implies
\begin{align*}
\phi_{r}^{(1)}(t) = \frac{1}{\omega_{r}}  \sum_{k=0}^{n} (-1)^{k} \left(    e^{i\omega_{r} t_{k+1}} - e^{i\omega_{r} t_{k}}     \right)  .  
\end{align*}
This leads to 
\begin{align*}
 \phi_{r}^{(1)}(t+\alpha) - \phi_{r}^{(1)}(\alpha) &= \\
 \frac{1}{\omega_{r}}  \sum_{k=0}^{n} (-1)^{k} &\left(    e^{i\omega_{r} (t)_{k+1}} - e^{i\omega_{r} (t)_{k}}     \right) \\
 &=\phi_{r}^{(1)}(t).
\end{align*}
Therefore, $\phi_{r}^{(2)}(t)$ is simplified as:
\begin{align*}
    \phi_{r}^{(2)}(t) = e^{-i\omega_{r}\alpha} \phi_{r}^{(1)}(t).
\end{align*}

\providecommand{\noopsort}[1]{}\providecommand{\singleletter}[1]{#1}%

\end{document}